\magnification1200
\centerline
{\bf  Conserved asymptotic charges for any massless particle }
\vskip 1cm
\centerline{\bf Kevin Nguyen  and  Peter West }
\vskip 1.2cm
\centerline{ {\it Department of Mathematics, King's College, London WC2R 2LS, UK}}
\vskip 1.7cm

\centerline{\sl Abstract}
We compute the conserved charges associated with the asymptotic symmetries of massless particles by examining their free theory in Minkowski spacetime. We give a procedure to systematically deduce the fall off of the massless fields at spatial infinity and show that it has a universal behaviour when expressed  in tangent space. We do this for generic massless particles. We do not impose gauge fixing conditions which allows us to uncover new nonzero charges for the graviton beyond the well-known supertranslation charges. We also compute conserved charges in the dual formulations of certain low spin particles and argue that this leads to an infinite number of new conserved charges.

\noindent
\vskip 6cm
emails: kevin.nguyen@kcl.ac.uk, peter.west540@gmail.com
\vfill

\eject

\medskip

{\bf 0. Introduction }
\medskip

Local symmetry transformations with support at infinity are associated with nonzero conserved charges. This subtle fact was first uncovered in the context of General Relativity through the search of a meaningful notion of energy and angular momentum for gravitating bodies [1-5]. An unexpected result was  the discovery by Bondi, van der Burg, Metzner and Sachs of an infinite number of  charges  associated with asymptotic symmetries which obeyed an algebra that was a generalisation of the Poincare algebra which included the so called supertranslations [6-8]. Much later a related result was found by Brown and Henneaux who showed  that in the context of three dimensional gravity with negative cosmological constant, that the asymptotic symmetries lead to conserved charges that satisfied an algebra that consisted of  two copies of the Virasoro algebra with non-zero central extension [9].  The revival of interest in the  subject of asymptotic symmetries  was due to the papers [10-11]. The former computed the charges of the supertranslations and their algebra, while the latter extended the asymptotic symmetries to include superrotations and calculated the corresponding algebra for supertranslations and superrotations. Since these  discoveries, the study of local asymptotic symmetries and corresponding charges has been extended in gravity [12-35] and applied to other gauge theories including electromagnetism [36-43], Yang-Mills [44-45] and supersymmetric theories [46-51]. In particular asymptotic symmetries associated with massless fields of arbitrary integer spins have been investigated in [52-54].
\par
The importance of asymptotic symmetries has become further apparent with the demonstration [55-56] that Ward identities for the  asymptotic symmetries have been shown to be equivalent to Weinberg's soft theorems in the scattering theory of quantum fields [57].   We refer the reader to [58] for a review of this interesting topic.
\par
In this paper we take the viewpoint that the asymptotic behaviour at spatial infinity  is encoded in the free theory in Minkowski spacetime. This is always possible as we are considering theories that are asymptotically flat and  the fields of the particles are small in this regime. We deduce the asymptotic fall off for the fields by demanding that they obey the equation of motions and the variation of the action is well defined. In practice this last requirement means that the boundary term which occurs when one varies the action vanishes. We can then compute the charges associated to any local symmetries at spatial infinity using the covariant phase space formalism originally developed in the papers  [9-10,59-60]. This task is greatly facilitated by working in tangent space, that is, tangential components of the fields share common fall off behaviour by contrast to the curvilinear coordinate components that are used throughout this work. Furthermore we do not  make a choice of gauge and so our results are independent of any gauge choice. This  also allows us to uncover new nonzero charges.
\par
Using this method we  will compute the conserved asymptotic charges for massless particles of spins  0,${}1\over 2$,1,${3\over 2}$ and 2.   When the fields are formulated with tangent indices we find a universal behaviour which is generically the same for all spins. The fields fall off goes as $r^{-1}$ and their  local gauge transformations as $r^{0}$ in four dimensions. Their conserved charges are non-zero and well defined. Indeed calculating in this way we can find the fall off at spatial infinity for  any massless particle and show that it has an infinite number of asymptotic charges. This is consistent with the fact that  soft theorems do exist in any theory containing massless particles.
\par
The above used the usual field descriptions of the particles. However, one can also use field descriptions that correspond to higher dualities and compute the conserved charges. Conserved charges associated with asymptotic symmetries have been computed in the dual formulations of certain  theories, as first discussed in the context of the electromagnetic duality in [61]. Rather than just work with the dual formulation we  write down parent actions that contain  both the usual field description and the dual descriptions of the particle concerned [62-65].  From these parent actions one can deduce  a description of the particle which contains both  the usual description or that in terms of the dual  field. In particular we carry this out for the spin zero particle for which the parent action contains the usual scalar field and the  dual two form field. See [66-70] for previous related work. We also compute the charges for the spin one particle  for the parent action for both the usual dual of the spin one particle as well as one of its higher dual formulations. This result indicates the presence of an infinite tower of new conserved charges associated with the infinite possible dual realisation of any particle.

\par
Before we begin we outline our conventions. We will be working with Minkowski space  in $D$ dimensions with the  coordinates $x^\mu=(t,r,\theta^m)$ and metric
$$
ds^2=g_{\mu\nu}\, dx^\mu dx^\nu=-dt^2+dr^2+r^2 \gamma_{mn}\, d\theta^m d\theta^n\,,
\eqno(0.1)$$
where $\theta^m, m=1,\ldots , D-2$ are arbitrary coordinates covering the  sphere $S^{D-2}$ and $\gamma_{mn}$ denotes the corresponding metric. It is related to the veirbein  $\hat{e}_m{}^\alpha$ on  the  sphere in the usual way
 $\gamma_{mn}=\hat e_m{}^\alpha \delta_{\alpha \beta} \hat e_m{}^\beta$. If we take spherical coordinates in four dimensions then $\theta^1=\theta$,  $\theta^2=\varphi$, and
$$
\hat {e}_m{}^\alpha=\big( \matrix  {1& 0\cr 0&\sin\theta \cr}\big)\,.
\eqno(0.2)$$
The nonzero Christoffel symbols in these coordinates are
$$
\Gamma^{r}_{mn}=-r \gamma_{mn}\,, \qquad \Gamma^m_{rn}=r^{-1} \delta^m_n\,, \qquad \Gamma^{m}_{nl}=\bar \Gamma^m_{nl}[\gamma]\,.
\eqno(0.3)$$
The vierbein  $e_\mu{}^a$ for the $D$  dimensional spacetime is related to the metric by $g_{\mu\nu}= e_\mu {}^a \eta_{ab} e_\nu{}^b$. It is  given in terms of the veirbein on the sphere by
$$
e_\mu{}^a= \big( \matrix  {1&0&0\cr 0&1&0\cr 0&0&r \hat e_m{}^\alpha \cr}\big)\,.
\eqno(0.3)$$

\medskip
{\bf 1. The massless spin zero particle}
\medskip
We begin by taking the usual description of a spin zero particle, namely a scalar field $\phi$ has the action
$$
S[\phi]=-{1\over 2}\int d^4x\, \sqrt{-\det g}\, \partial^\mu \phi\, \partial_\mu \phi\,.
\eqno(1.1)$$
where in the above coordinates $\sqrt {-\det g}= r^2\det \hat e$ while   $d^4x= dtdrd\theta ^m$.
An arbitrary variation of the action gives
$$
\delta S=\int d^4x\, \delta\phi \partial_\mu ( \sqrt{-\det g} \partial^\mu \phi) -\int d^4x \partial_\mu (\sqrt{-\det g}\delta \phi \partial^\mu \phi)\,.
\eqno(1.2)$$
From the first term we read off the equation of motion to be
$$
{1\over  \sqrt{-\det g}} \partial_\mu ( \sqrt{-\det g} \partial^\mu \phi) = -r^2 \ddot \phi +\partial_r(r^2\partial_r \phi) + {1\over (\det \hat e ) }\partial_m ((\det \hat e )\gamma^{mn} \partial_n \phi)=0\,.
\eqno(1.3)$$
\par
We are interested in the asymptotic behaviour of the field and we take it to be of the form
 $$
\phi(r,t,\theta^m)=\phi^{(0)}(t, \theta^m)+{\phi^{(1)}(t, \theta^m)\over r}+{\phi^{(2)}(t, \theta^m)\over r^2}+O(r^{-3})\,.
\eqno(1.4)$$
which is the most general expansion such that $\phi$ is  finite  as $r\to \infty$ provided we assume that there is no logarithmic behaviour.
\par
We will now determine the asymptotic behaviour of $\phi$ by demanding that it obey the equation of motion and the boundary term on equation (1.2) vanishes. This term takes the form
$$
-\int_{S^2} dt\, d\Omega\, r^2   \delta \phi \partial_r \phi =  \int_{S^2}  dt\, d\Omega \delta \phi^{(0)}\,  \phi^{(1)}+O(r^{-1})\,,
\eqno(1.5)$$
where the integration is over the sphere $S^2$ at infinity and $d\Omega= d^2\theta^m \det \hat e$. To have a well defined variation we require that this vanishes as $r\to \infty$ for all variations and so we require
$$\delta \phi^{(0)}\,  \phi^{(1)}=0\,.
\eqno(1.6)$$
Hence either $\phi^{(0)}$ or  $\phi^{(1)}$ is zero.
\par
Substituting the asymptotic expansion of equation (1.4) into the field equation (1.3) we find that
$$
\ddot \phi^{(0)}=0= \ddot \phi^{(1)}=-\ddot \phi^{(2)}+ {1\over (\det \hat e ) }\partial_m ((\det \hat e )\gamma^{mn} \partial_n \phi^{(0)}), \ldots
\eqno(1.7)$$
The solution to equations (1.6) and (1.7) is given by
$$
 \phi^{(0)}=0 , \ \dot  \phi^{(1)}=0 ,\  \ddot  \phi^{(2)}=0\,,
\eqno(1.8)$$
which implies that $\phi$ has the expansion
$$
\phi= {\phi^{(1)}(\theta^m)\over r}+{\phi^{(2)}(\theta^m)\over r^2}+O(r^{-3})\,.
\eqno(1.9)$$
In this last step we discarded the possibility that   $\phi^{(2)}$ can have a linear growth in  time.
\par
We will now discuss the constraints imposed on the asymptotic expansion which are due to Poincare symmetry.
Poincare transformations are generated by the vector fields which in  Cartesian coordinates take the form
$$
\xi^t=b_i x^i+a, \qquad \xi^i=b^i t+b^i_{\ j} x^j+a^i, \qquad b_{ij}=-b_{ji}.
\eqno(1.10)$$
While, in spherical coordinates, they are given by [40,70]
$$
\xi^t=b r+a, \qquad \xi^r=W+b t, \qquad \xi^m=Y^m+r^{-1} \partial^m (W+b t),
\eqno(1.11)$$
where $b,W$ are linear combinations of the $l=1$ spherical harmonics and respectively parameterise boosts and spatial translations. Also $Y^m$ are the three rotations on the sphere, that is, the three Killing vectors.
\par
Acting on the asymptotic form of equation (1.4) of the scalar field the boost  of equation (1.11) leads to the result
$$
\delta_\xi \phi=\xi^\mu \partial_\mu \phi=b\, \dot{\phi}^{(1)}+O(r^{-1})\,,
\eqno(1.12)$$
This will only preserve the asymptotic expansion if $\dot{\phi}^{(1)}=0$. In this way we recover our previous result but in a more straightforward way.
\par
Even though the scalar, as formulated above has no apparent symmetries, it has an infinite number of conserved charges which are given by [70]
$$
Q[\lambda]=\int_{S^2} d\Omega\, \lambda(\theta^m)\, \phi^{(1)}(\theta^m)\,.
\eqno(1.13)$$
The charges are parameterised by $\lambda(\theta^m)$ and it  are conserved due to equation (1.8).
\medskip
A spin zero particle can also be described by a two form field $A_{\mu\nu}$. A parent action which contains the above and the two form formulation of a spin zero particle is given by
$$
S^D=\int d^4 x (-{1\over 2} \sqrt{-\det g} F_\mu F^\mu - {1\over 3!} \epsilon ^{\mu\nu_1\nu_2\nu_3} F_{\mu}\partial_{\nu_1}A_{\nu_2\nu_3})\,,
\eqno(1.14)$$
where we have also introduced the one form field $F_\mu$. An arbitrary variation of this action gives
$$
\delta S^D=\int d^4x\, (-\delta F_\mu ( \sqrt{-\det g} F^\mu - {1\over 3!} \epsilon ^{\mu\nu_1\nu_2\nu_3} \partial_{\nu_1} A_{\nu_2\nu_3})
+{1\over 3!} \delta A_{\nu_2\nu_3} \epsilon ^{\mu\nu_1\nu_2\nu_3}\partial_{\nu_1}  F_{\mu})
$$
$$
 - {1\over 3!} \int d^4x \partial_{\nu_1} (\epsilon^{\mu\nu_1 \nu_2\nu_3} F_\mu  \delta A_{\nu_2\nu_3})\,.
 \eqno(1.15)$$
\par
The equations of motion can be read off from equation (1.15) are
$$
\sqrt{-\det g} F^\mu +{1\over 3!} \epsilon ^{\mu\nu_1\nu_2\nu_3} \partial_{\nu_1} A_{\nu_2\nu_3} =0 ,\ \
\partial_{[\nu}  F_{\mu ]}=0\,.
\eqno(1.16)$$
The second equation implies that $F_{\mu }=\partial_{\mu }\phi$ and substituting this back in the action we recover the action of equation (1.1). Alternatively  substituting for $F_\mu$ as given in the first  equation,  we find the action
$$
-{1\over 2.  3!}\int d^4 x \sqrt {-\det g}G_{\mu_1\mu_2\mu_2} G^{\mu_1\mu_2\mu_3}\,,
\eqno(1.17)$$
where $G_{\mu_1\mu_2\mu_2}= \partial_{[\mu_1} A_{\mu_2\mu_3]}$. Thus we find an alternative description of a spin zero particle.
\par
We will now derive the asymptotic behaviour in this theory by demanding that  the equations of motion hold and that the boundary term should vanish. We will also derive the conserved charges. We will do this in detail using the usual formulation but at the end of this section we will work in tangent space and re-derive these results with a fraction of the effort. The reader who wants to get ahead can skip to the end of this section.
\par
The boundary condition of equation (1.15) takes the form
$$
- {1\over 3!} \int dt d^2\theta \epsilon^{r \mu\nu_1\nu_2} F_\mu  \delta A_{\nu_1\nu_2}
\eqno(1.18)$$
as $r\to \infty$. Since the variations are arbitrary each of the above terms must vanish, we require that
$$
F^{(1)}_t  \delta A^{(-1)}_{mn}= 0=F^{(0)}_t  \delta A^{(-1)}_{mn} =F^{(0)}_t  \delta A^{(0)}_{mn}   ,
\eqno(1.19)$$
$$
F^{(1)}_{n} \delta A^{(-1)}_{m t}= 0\,,              
\eqno(1.20)$$
and
$$
F^{(0)}_{ n} \delta A^{(-1)}_{m t} = 0    = F^{(0)}_{n}  \delta A^{(0)}_{m t}\,.
\eqno(1.21)$$
\par
We will now determine the asymptotic behaviour of the fields by demanding that it obeys the equations of motion and the boundary term vanishes. The fall off can be written as
$$
A_{\mu\nu}= A_{\mu\nu}^{(-1)}(t,\theta^n)r + A_{\mu\nu}^{(0)}(t,\theta^n)+ {A_{\mu\nu}^{(1)}(t,\theta^n)\over r}+O(r^{-2}),\
$$
$$
F_\mu (r,t,\theta^m)=F_\mu^{(0)}(t,\theta^m)+{F_\mu^{(1)}(t,\theta^m)\over r}+{F_\mu^{(2)}(t,\theta^m)\over r^2}+O(r^{-3}) , \ \
$$
$$
\phi(r,t,\theta^m)=\phi^{(0)}(t,\theta^m)+{\phi^{(1)}(t,\theta^m)\over r}+{\phi^{(2)}(t,\theta^m)\over r^2}+O(r^{-3})\,.
\eqno(1.22)$$
We have also given the expansion of $\phi$ in terms of which  $F_\mu$ is determined.
\par
It is straightforward to show that the equation of motion (1.3) implies that
$$
F^{(0)}_r=0= F^{(1)}_r= F_t^{(0)}=F^{(1)}_t\,.
\eqno(1.23)$$
The last equation is shown by first showing $\dot F_t^{(1)}=\ddot \phi^{(1)}=0$ and then ruling out any linear dependence on time to conclude that $F_t^{(1)}=0$. Thus equations (1.19) are satisfied. The equations of motion can then be used to show that
$$
A^{(-1)}_{nr}=\partial _n \zeta^{(-1)}(t,\theta^n), \ A^{(-1)}_{tr}= \dot \zeta^{(-1)}(t,\theta^n),
\eqno(1.24)$$
which implies that there are  of the form of gauge transformations which we can set to zero.  Since the field $A_{\mu\nu}$ appears in its field strength it is inevitable we will find gauge transformations in the asymptotic analysis. Thus we have
$$
A^{(-1)}_{nr}=0=A^{(-1)}_{tr}.
\eqno(1.25)$$
\par
To avoid the possibility that $\phi^{(1)}$ is a constant
we satisfy equation (1.20) by taking
$$
A^{(-1)}_{nt}=0,
\eqno(1.26)$$
which, using the equations of motion, implies  that $\dot A^{(-1)}_{nm}=0$.
\par
To avoid the possibility that $\phi^{(0)}$ is non zero we take $F_m^{(0)}=0$ in order to satisfy equation (1.21). Whereupon,
using the equations of motion,  one can show that $\dot F^{(2)}_t=\ddot \phi^{(2)}=0$ and, using the same argument as above,  we take $F^{(2)}_t=0$. Using this result, the equations of motion can be used to show that
$$
A^{(0)}_{nr}=\partial _n \zeta^{(0)}(t,\theta^n) + a^{(0)}_{nr}(\theta^n), \ A^{(0)}_{tr}= \dot \zeta^{(0)}(t,\theta^n).
\eqno(1.25)$$
We recognise the terms involving $\zeta^{(0)}$ as a gauge transformation  which we can remove by a gauge choice and so we take $A^{(0)}_{tr}= 0$. This leaves the gauge transformation with parameter $\zeta^{(0)}$  which is independent of time.
\par
Thus finally we arrive at the asymptotic expansion for our fields
$$
A_{mn}=A_{mn}^{(-1)}(\theta^m)r + O(r^0) , \ \ A_{mr}= A_{mr}^{(0)} (x^n)  + O(r^{-1}) ,
$$
$$
A_{tr}= A_{tr}^{(1)}(t, \theta^m)r^{-1} + O(r^{-2}) ,\ \
A_{tm}= A_{tn}^{(0)}(t, \theta^m)r^0 + O(r^{-1}) ,\
\eqno(1.26)$$
as well as the expansion of $\phi$ of equation (1.9). We note that as long as the field has no component in the time direction then the lead term is independent of time.
\par
The action of equation (1.14) has the obvious gauge symmetry
$$
\delta A_{\mu\nu}= \partial_{[\mu} \Lambda_{\nu]}.
\eqno(1.29)$$
This is consistent with the asymptotic expansion of the fields of equation (1.2) provided we take the gauge parameter $\Lambda_\mu$ to have the fall off
$$
\Lambda_m= \Lambda^{(-1)}_ m (\theta^m) r+ O(r^0) ,\ \ \Lambda_r= \Lambda^{(0)}_r (t, x^n) r^0+ O(r^{-1}) ,\ \
\Lambda_t= \Lambda^{(0)}_t (\theta^m) r^0+ O(r^{-1}).
\eqno(1.30)$$
\par
We will now compute the  conserved charges arising from the above gauge transformation. The symplectic potential $ \theta$ is read off from the boundary term of equation (1.15) to be
$$
 \theta[\delta A]={1\over 3!} \int_\Sigma d^3x \epsilon^{0\mu\nu\nu} F_\mu \delta A_{\nu_1\nu_2},
\eqno(1.31)$$
where $\Sigma$ is a time slice of spacetime. The variation of the associated conserved charge $Q_\Lambda$  is given by
$$
\delta Q_\Lambda = \delta_\Lambda  \theta[\delta A] - \delta  \theta[\delta_\Lambda A].
\eqno(1.32)$$
One finds that
$$
Q_\Lambda = -{1\over 3!} \epsilon^{0rmn}\int_{S^2} d^2\theta F_m \Lambda_n=  \int_{S^2} d^2\theta \phi \lambda,
\eqno(1.33)$$
where the integral is over the sphere at infinity and $ \lambda= {1\over 36} \epsilon^{0rmn}\partial_m \Lambda_n $. It is easy to see that the charge is finite and non-zero as $F_m\sim r^{-1}$ while $\Lambda_m\sim r^{1}$. Thus from the parent action we recover the conserved charges of equation (1.13) but they now arise from a symmetry of the action. The asymptotic scalar charge of equation (1.33) was observed to be present in reference [66] and was computed from the dual scalar formulation in reference [67]. However, using the parent action we see that the same charge is indeed related to both ways of formulating a spin zero particle.
\par
For the important example of a massless scalar field, it was quickly realised that the corresponding soft theorem can indeed be viewed as the conservation law of an asymptotic charge, even though a Noetherian interpretation in terms of an asymptotic local symmetry was initially missing [66]. A connection to gauge symmetries could finally be achieved by going to a formulation of the massless spin zero particle  in terms of a dual two-form gauge field [67]. Thus the existence of a relation between asymptotic charges and asymptotic gauge symmetries depends on the chosen formulation of the theory.
\par
We will now write the above results in tangent space. Using the vierbein in equation (0.3) we find that the asymptotic fall off of the fields of equation (1.28) and the parameter of equation (1.30) take the form
$$
A_{ab}= {A^{(1)}_{ab}(\theta^m)\over r} +O(r^{-2}) ,\ \ \Lambda_a  = \Lambda^{(0)}_a (\theta^m) r^0+O(r^{-1}).
\eqno(1.34)$$
We note that in tangent space the fall off is the same for all components.
\par
The derivative $\nabla_a= (e^{-1})_a{}^\mu \partial_\mu $ is of order $r^{-1}$ unless it  involves a derivative with respect to $t$. However the leading terms in the asymptotic expansion are independent of time and and so if the derivative involves a time derivative  it will only be non-zero on the next to leading term and so in effect we have a $r^{-1}$ factor. Hence  in effect the derivative $\nabla_a$ always contributes a factor of  $r^{-1}$. We note that applying the derivative to  $\Lambda_a$ to find the gauge transformation of   $A_{ab}$ gives the correct $r $ dependence.
\par
Using tangent space from the outset considerably simplifies the calculation. Since this is the way we will treat the other theories in tis paper we will now explain how it goes for the above theory. The equations of motion (1.16) in tangent space takes the form
$$
\nabla_{[a} F_{b]}=0,\ \ F_a+ {1\over 3!} \epsilon _a{}^{b_1b_2b_3} \nabla_{b_1} A_{b_2b_3}=0.
\eqno(1.35)$$
While the boundary term of equation (1.18) takes the form
$$
-{1\over 3!} \int dt d\theta^2 \epsilon ^{ra_1a_2a_3} F_{a_1} \delta A_{a_2a_3} \det  e.
\eqno(1.36)$$
We recall that $\det e= r^2 \det \hat e$. Since the equation of motion involves no factors of $r$ and the boundary term only involves an overall factor of $r^2$, all the components of the fields, when written in tangent space,  occur in the same way. As the asymptotic behaviour is determined by the field equations and demanding that the boundary term vanish it follows that  all components of a given field in tangent space   must have the same asymptotic expansion. If the leading behaviour of $A_{a_2a_3} $ is of order $r^{-n}$ then, by equation (1.35),  $F_a$ must have the leading behaviour  $r^{-n-1}$ as the derivative goes as $r^{-1}$.
Demanding that the boundary term of equation (1.36)  vanish for any variation requires that
$F_{a_1}\delta  A_{a_2a_3}$ must have the leading term $r^{_3}=r^{-2n-1} $ and so we conclude that $n=1$. As the derivative in tangent space introduces a $r^{-1}$ factor we conclude that the gauge parameter in tangent space must have leading behaviour $r^0$. The charge of equation (1.33) has the form
$$
Q_\Lambda = -{1\over 3!} \epsilon^{0r\alpha\beta }\int_{S^2} d^2\theta  r^2  \det \hat e F_\alpha \Lambda_\beta,
\eqno(1.37)$$
and so it is well behaved and non-zero as $r\to \infty$. As such we have quickly recovered the asymptotic behaviour of the spin zero in its dual description. The tangent space  treatment for the spin zero in its usual description recovers the asymptotic analysis at the beginning of this section in a few lines.
\medskip
{\bf 2. Spin one }
\medskip
Let us begin with the formulation of spin one due to Maxwell and so we take the action
$$
S^M=-{1\over 4} \int d^4 x F_{\mu\nu} F^{\mu\nu}\sqrt{-\det g},
\eqno(2.1)$$
where $F_{\mu\nu}=\partial_{[\mu}A_{\nu]}$ and we have the usual  gauge symmetry $\delta A_\mu=\partial_\mu \Lambda$. The variation of the action is
$$
\delta S^M= \int d^4 x\delta A_\nu \partial_\mu(\sqrt{-\det g}F^{\mu\nu})
- \int d^4 x\partial_\mu (\sqrt{-\det g} \delta A _\nu F^{\mu\nu}).
\eqno(2.2)$$
The equation of motion is
$$
\partial_\mu(\sqrt{-\det g}F^{\mu\nu})=0,\ \ {\rm or \ in \ tangent \ space}\ \ \nabla_a F^{ab}=0,
\eqno(2.3)$$
and the boundary term is
$$
- \int_\Sigma  dt d\Omega r^2 \delta A _\nu F^{r\nu}) ,\ \ {\rm or\  in\  tangent \ space}\ \ - \int_\Sigma  dt d\Omega r^2 \delta A _a F^{ra}).
\eqno(2.4)$$
\par
Using equation (1.32) we find the symplectic potential and as a result the conserved charges corresponding to the gauge symmetry. These quantities are
$$
 \theta [\delta A_\nu] =- \int dr d^2 \theta  \delta A_\nu F^{0\nu} \sqrt{-\det g},\ \ Q(\Lambda )=\int d\Omega r^2 F_{0r} \Lambda (\theta^m)
\eqno(2.5)$$
\par
It is simple to compute the asymptotic behaviour of the fields  as long as we consider the fields in  tangent space. If $A_a$ has the lead behaviour $r^{-n}$ as $r\to \infty$. Then $F_{ab}$ behaves as $r^{-n-1}$ and $\delta A _a F^{ra}$ as $r^{-2n-1}$ . For the boundary term to vanish we require $\delta A_a F^{ra}$ to go as $r^{-3}$ and so $n=1$. As such we have
$$
A_{a}= {A^{(1)} _{a}(\theta^m) \over r} +O (r^{-2}), \ \ \Lambda = \Lambda^{(0)}(\theta^m)+ O(r^{-1}).
\eqno(2.6)$$
The fall off for the potential with world indices is easily found given the vielbein. The absence of a time dependence in the leading terms will be discussed later in the paper.
It follows that the charges of equation (2.5) are well defined and non-zero as $r\to \infty$.
\par
We now consider the parent action
$$
S^{MD}= \int d^4 x ( -{1\over 4} F_{\mu\nu} F^{\mu\nu}\sqrt{-\det g}+{1\over 4} \epsilon^{\mu \nu \rho \kappa} F_{\mu\nu} \partial_\rho B_{\kappa}),
\eqno(2.7)$$
where $F_{\mu\nu}$ is an independent field and $B_\mu$ is subject to the gauge transformation  $\delta B_\mu=\partial_\mu \Lambda^D$. The field equation for $B_\mu$ implies the Bianchi identity for $F_{\mu\nu}$ and so
$F_{\mu\nu}= \partial _{[\mu} A_{\nu]}$ and using this we recover our first action. On the other hand  we can use the  field equation for $F_{\mu\nu}$  to eliminate $F_{\mu\nu}$ to find an action for the dual field $B_{\mu}$.
\par
The variation of the action, when written in tangent space,  is given by
$$
\delta S^{MD}=-{1\over 2} \int d^4x  \sqrt{-\det g} \delta F_{ab}(F^{ab}-{1\over 2} \epsilon^{ab cd} \nabla_c B_{d}) -{1\over 4} \int d^4 x \delta B_{d}  \epsilon^{ab c d} \nabla_{c}  F_{ab}
$$
$$
+{1\over 4} \int dt d^2\theta  \det e  \epsilon^{rabc} \delta F_{ab}B_c.
\eqno(2.8)$$
The conserved charge associated with the gauge transformation $\Lambda ^D$ computed using equation (1.32) is given by
$$
Q(\Lambda^D) = {1\over 4} \int d^2\theta \epsilon ^{0rmn} F_{nm}\Lambda ^D = {1\over 4} \int d\Omega  r^2 \epsilon^{\alpha\beta } F_{\alpha \beta }\Lambda ^D,
\eqno(2.9)$$
where in the last equation we have put the expression in tangent space. We recognise the magnetic charge.
\par
We will now deduce the asymptotic behaviour. Let us suppose $B_a$ has the leading behaviour $r^{-n}$ then from the equation of motion $F_{ab}$ goes as $r^{-n-1}$, $\Lambda^D$ as $r^0$  and so  $\delta F_{ab}B_c$ goes as $r^{-2n-1}$. Demanding the boundary term is well defined implies that $n=1$. It follows that the charge of equation (2.9) is finite as $r\to \infty $ and is non-zero.
\par
We now consider an alternative dual formulation of a spin one particle. Rather than represent the particle by $A_\mu$ or $B_\mu$, we can take the field $A_{\mu\nu |\lambda}$ where $A_{\mu\nu |\lambda}=-A_{\nu\mu |\lambda}$  but is otherwise unrestricted. A corresponding parent action is given by
$$
S^{MDD}=\int d^4 x   \sqrt{-\det g}(-\epsilon ^{a b_1b_2b_3}P_{a |}{}^c \nabla_{b_1}A_{ b_2b_3 | c} -{1\over 2}P_{a | b} P^{a | b} +{1\over 2} P_{a |}{}^{a} P_{b |}{}^{b})).
\eqno(2.10)$$
It has the gauge symmetries
$$
\delta P_{\mu|\nu}= D_\mu\partial_\nu \Lambda ,\ \ \delta A_{\mu\nu|\tau }= -{1\over 2} \epsilon _{\mu\nu \tau}{}^\kappa \partial_\kappa \Lambda , \ \ \ \ \delta A_{\mu\nu|\tau }=D_{ [\mu}\lambda_{\nu] |\tau}.
\eqno(2.11)$$
The variation of the action is given by
$$
\delta S^{MDD}=\int d^4 x \sqrt{-\det g} \delta P_{a | c}(-\epsilon ^{a b_1b_2b_3}\nabla_{b_1}A_{ b_2b_3 | c} - P^{a | c} +{1\over 2} \eta^{ac}   P_{b |}{}^{b})
$$
$$
+\int d^4 x \epsilon ^{a b_1b_2b_3} \nabla_{b_1} (\sqrt{-\det g} P_{a |}{}^c) \delta A_{ b_2b_3 |}{}_{ c}
+\int _\Sigma d^2 \theta dr \sqrt{\det g} \epsilon ^{0 ab_1b_2 }\delta A_{b_1b_2 | c} P_{a |}{}^{c}.
\eqno(2.12)$$
\par
From the $A_{b_1b_2|c}$ equation of motion we find that $P_{a|b}= \nabla _a A_b$ and substituting this into the action we find the usual action for a spin one particle. On the other hand  eliminating $P_{a|b}$ from its equation of motion we find the action
$$
\int d^4 x \sqrt{-\det g}  ( -3 G_{a_1a_2a_3 | b}G^{a_1a_2a_3 | b}+4 G_{[a_1a_2a_3 | b]}G^{[a_1a_2a_3 | b]}),
\eqno(2.13)$$
where $G_{a_1a_2a_3 | b}= \nabla _{[a_1 }A_{a_2a_3]|b}$.
\par
The derivation of the asymptotic behaviour of the fields follows the same pattern as our early theories. If $A_{a_1a_2 | b}$
has the leading behaviour $r^{-n}$ then $P_{a|b}$ goes as $r^{-n-1}$. Then $\delta A_{b_1b_2 | c} P_{a |}{}^{c}$ goes as
$r^{-2n-1}$ and requiring that  the boundary term to vanish we must require that  this expression  must go as $r^{-3}$  and so $n=1$. The parameters $\Lambda$ and $\Lambda _{a | b}$ must both  go as $r^0$.
\par
Using equation (1.32) we find the conserved charges are
$$
Q= \int d^2 \theta dr \det e \epsilon^{0a b_1 b_2} (\nabla_a \nabla^c \Lambda A_{b_1b_2 | c} +{1\over 2} \epsilon_{b_1b_2 c d} P_{a |}{}^c \nabla ^d\Lambda -P_{a |}{}^c \nabla _{b_1} \lambda_{b_2 |c} ).
\eqno(2.14)$$
Using the duality relations we may write this as
$$
Q=\int d\Omega r^2 (f^{0r}\Lambda + \epsilon^{\alpha\beta}f_{\alpha} {}^\gamma \lambda_{\beta | \gamma} )
+\int d\Omega r^2 \epsilon^{\alpha\beta} (\nabla^c\Lambda A_{\alpha\beta |c} +\nabla^c A_{\alpha} \lambda_{\beta |c}
+f_{\alpha} {}^0 \lambda_{\beta | 0}+f_{\alpha} {}^r \lambda_{\beta | r}),
\eqno(2.15)$$
where $f_{\mu\nu}= \partial_\mu A_\nu -\partial_\nu A_\mu$. With the boundary conditions we have taken above all  the terms in the charge are finite and non-zero as $r\to \infty$.
\par
 In the first term of equation (2.15) we recognise the usual electric and magnetic charges and so  the parent action of equation (2.10) contains both of these charges. One might impose stronger boundary conditions on $A_{a_1a_2,b}$,  and so on the parameter $\Lambda_{a|b}$,  such that the additional contributions go away. However, it would seem that the higher dual formulations lead to new unfamiliar charges whose  physical significance we hope to elucidate in a later paper.

\medskip
{\bf 3. n forms and their duals in D dimensions }
\medskip
We will now find the asymptotic behaviour and the conserved charges for a n form field in $D$ dimensions that has the action
$$
S^n= -\int d^D x {1\over 2 (n+1)}\sqrt {-\det g} F_{\mu_1\ldots \mu_{n+1}}F^{\mu_1\ldots \mu_{n+1}},
\eqno(3.1)$$
where $ F_{\mu_1\ldots \mu_{n+1}}=(n+1) \partial _{[\mu_1}A_{\mu_2\ldots \mu_{n+1}]}$ and we have the obvious gauge symmetry\break
$\delta A_{\mu_1\ldots \mu_{n}}= n\partial_{[\mu_1}\Lambda _{\mu_2 \ldots \mu_{n]}}$. See [71] for previous work. The variation of the action is given by
$$
\delta S^n= \int d^Dx \delta A_{\mu_2\ldots \mu_{n+1}}\partial_{\mu_1} ( \sqrt {-\det g}  F^{\mu_1\ldots \mu_{n+1}})
-\int dtd\Omega  r^{D-2} \delta A_{a_1\ldots a _{n}}  F^{ra_1\ldots a_{n}}.
\eqno(3.2)$$
The sympletic potential can be read off to be
$$
 \theta [\delta A_{\mu_1\ldots \mu _{n}}  ]=-\int_\Sigma dr d^{D-2}\theta \delta A_{\mu_1\ldots \mu _{n}}  F^{0 \mu_1\ldots \mu_{n}}\sqrt {-\det g}.
\eqno(3.3)$$
Using equation (1.32) we find that the  conserved charges associated with the gauge symmetry are given by
$$
Q(\Lambda _{a_1\ldots a_{n-1}} )= \int _{S^{D-2}} d\Omega r^{D-2} \Lambda _{ \alpha_1\ldots \alpha _{n-1}} F^{0r \alpha_1\ldots \alpha_{n-1}}.
\eqno(3.4)$$
\par
If we take $A_{a_1\ldots a_{n}}$ to have the leading behaviour $r^{-n}$ as $r\to \infty$  then $F_{a_1\ldots a_{n+1}}$ goes
$r^{-n-1}$ and the parameter $\Lambda _{\mu_1 \ldots \mu_{n-1}}$ as $r^{-n+1}$ . Demanding that the boundary term vanish at infinity  implies that $n={1\over 2}(D-2)$. One finds that the conserved charges are well defined and non-zero.
\par
The dual field to $A_{\mu_1\ldots \mu_{n}}$ is a $m=D-n-2$ rank gauge field $B_{\mu_1\ldots \mu_{m}}$. A parent action is given by
$$
S^{Dm}= {1\over (n+1)!}\int d^D x (-{1\over 2 m!}\sqrt {-\det g} F_{\mu_1\ldots \mu_{n+1}}F^{\mu_1\ldots \mu_{n+1}}
$$
$$
+{1\over (m+1)!}\epsilon^{\mu_1\ldots \mu_{n+1} \nu_1\ldots \nu_{m+1}} F_{\mu_1\ldots \mu_{n+1}}\partial_{\nu_1}B_{\nu_2\ldots \mu_{m+1}}).
\eqno(3.5)$$
The $B_{\mu_1\ldots \mu_{m}}$ field equation of motion implies that $F_{\mu_1\ldots \mu_{n+1}}= (n+1)! m! \partial_{[\mu_1}A_{\ldots \mu_{n+1}]}$ and substituting this into the action we find the original action of equation (3.1). Alternatively eliminating the field $F_{\mu_1\ldots \mu_{n+1}}$ using its equation of motion we find an analogous action but for $B_{\mu_1\ldots \mu_{m}}$.
\par
Varying the action we find the boundary term
$$
{1\over (m+1)!}{1\over (n+1)!}\int dt d\Omega r^{D-2}\epsilon ^{a_1\ldots a_{n+1}rb_1\ldots b_{m}}F_{a_1\ldots a_{n+1}}\delta B_{b_1\ldots b_{m}},
\eqno(3.6)$$
and the conserved charges
$$
Q(\Lambda _{a_1\ldots a_{m-1}} )= -{1\over (m+1)!}{1\over (n+1)!}\int  d\Omega r^{D-2}\epsilon ^{\alpha_1\ldots \alpha_{n+1}\beta_1\ldots \beta _{m-1}}F_{\alpha_1\ldots \alpha_{n+1}}\Lambda_{\beta_1\ldots \beta_{m-1}}.
\eqno(3.7)$$
\par
Using the same arguments as above we find that the gauge fields $B_{b_1\ldots b_{m}}$ and $A_{b_1\ldots b_{n}}$ both have the leading behaviour $r^{-{(D-2)\over 2}}$, while $F_{a_1\ldots a_{n+1}}$ goes as $r^{-{D\over 2}}$ and the gauge parameter as $\Lambda _{a_1\ldots a_{m-1}}$ as $r^{-{(D-4)\over 2}}$. Thus we find the magnetic charge as opposed to the electric charge.
\par


Form fields are more usually associated with brane charges and so we will now consider how asymptotic charges fit into this case. A rank $n\equiv p+1$ gauge field naturally provides a charge for  a $p$ brane. Following the often adopted convention we take the coordinates of spacetime to be such that the brane sits in the directions $x^\mu , \mu=0,1,\ldots p$ and the coordinates transverse to the brane to be given by  $x^{\mu^\prime} , \mu^\prime =p+1,\ldots D-1$. Corresponding to the presence of the brane  we take 
the line element to be of the form 
$$
d^2s= -dt^2 + \sum_{\mu=1}^{p} d^2 x_\mu+ d^2r + r^{D-p-2} \sum_{m^\prime=1}^{D-p-2}d\theta ^{m^\prime }\gamma^p_{m^\prime n^\prime}d\theta ^{n^\prime}
\eqno(3.8)$$
where $\gamma^p_{m^\prime n^\prime}$ is the metric on the sphere $S^{D-p-2}$  in the spatial coordinates transverse to the brane which has the coordinates $ \theta ^{m^\prime}, \ m^\prime=1,\ldots , D-p-2$ with vierbein $e_{m^\prime}{}^{\alpha^\prime}$. 
\par
The integral over all of space and time then splits into a part over the spatial volume of the brane, the time and the spatial coordinates transverse to the brane. As the brane has translation symmetry the integral over its spatial world volume just gives  the spatial volume of the brane, denoted $V_p$. The action of the brane which replaces that of equation (3.1)  then becomes  
$$
S^n= -V_p\int dt d^{D-p-2} x {1\over 2 (n+1)}\sqrt {-\det g}  F_{ \mu_1\ldots  \mu_{p+2}}F^{ \mu_1\ldots  \mu_{p+2}},
\eqno(3.9)$$ 
where the indices $  \mu$ take the range $  \mu=p+2,\ldots , D-1$. 
\par
Varying the action we find the boundary term 
$$
B_p=V_p\ \int dt dr d\Omega_p r^{D-p-2} F^{r  \mu _1\ldots   \mu_{p+1}} \delta A_{  \mu_1\ldots   \mu_{p+1}}
\eqno(3.10)$$
where $d\Omega_p= d^{D-p-2} \theta^{m^\prime} \sqrt {\det \gamma^p_{m^\prime n^\prime}}$ for $m^\prime=1,\ldots , D-p-2$. For this boundary term to vanish we require 
$$
A_{  \mu_1\ldots   \mu_{p+1} }= {A^{(1)}{}_{  \mu_1\ldots   \mu_{p+1}}(\theta^m) \over r^{D-p-2\over 2}}+\ldots ,\ \ 
\Lambda_{  \mu_1\ldots   \mu_{p} }= {\Lambda^{(0)}{}_{  \mu_1\ldots   \mu_p}(\theta^m) \over r^{D-p-4\over 2}}+\ldots  
\eqno(3.11)$$
\par
The symplectic potential is given by 
$$
 \theta [\delta A_ {\mu_1\ldots \mu _{n} }  ]=-\int_\Sigma dr d\Omega_p r^{D-p-2} \delta A_{  \mu_1\ldots   \mu _{n}}  F^{0   \mu_1\ldots   \mu_{n}}. 
\eqno(3.12)$$
and we find that the conserved asymptotic  charges are  
$$
Q_{\Lambda}  =V_p\ \int _{S^{D-p-2}} d\Omega_p r^{D-p-2} \Lambda _{ \alpha_1\ldots \alpha _{p}} F^{0r \alpha_1\ldots \alpha_{p}}.
\eqno(3.13)$$
We recognise from equation (3.11) that it is well defined and non-zero. 
\par
It is instructive to compute the charge for the M2 brane solution for which $F_{012r}= \partial_r N^{-1}$ where $N=1+{|Q|\over 6r^6}$. The charge is 
$$
Q_{\Lambda}  =2V_2\ \int _{S^{7}} d\Omega_2 r^{7} \Lambda _{ 12} F^{0r 12}=2V_2 |Q| \int _{s^7} d\Omega_2 \Lambda _{12}
\eqno(3.14)$$
The spherical nature of the brane in the transverse directions implies that only one charge is non-zero. 
\par
One can also apply the asymptotic analysis to the world volume fields on the brane. This follows the above discussions except that it is confined to the brane world volume. 


\medskip
{\bf 4. Gravity in D dimensions}
\medskip
The action for the free spin two particle in $D$ dimensions  is given by
$$
S^{(2)}= \int d^D x (-\nabla_\tau h^{\mu\nu } \nabla^\tau h_{\mu\nu} +2 \nabla_\mu  h^{\mu\nu} \nabla_\tau h^\tau{}_\nu -2 \nabla_\mu h^{\mu\nu }\nabla_{\nu} h^\tau{}_\tau+\nabla_\mu h^\nu{}_\nu \nabla^\mu h^\tau{}_\tau )\sqrt {-det g}.
\eqno(4.1)$$
Varying the action we find that
$$
\delta S^{(2)}=2\int d^D x \delta h^{\mu\nu} E_{\mu\nu} \sqrt {-det g}+  B,
\eqno(4.2)$$
where $E_{\mu\nu} $ is the equation of motion and $B$ is the boundary term. These are given by
$$
E_{\mu\nu} \equiv  \nabla^2 h_{\mu\nu} - \nabla_\mu \nabla_\tau h^\tau{}_\nu-\nabla_\nu \nabla_\tau h^\tau{}_\mu+\nabla_\mu \nabla_\nu h^\tau{}_\tau-g_{\mu\nu} (\nabla^2 h^\tau{}_\tau - \nabla_\tau \nabla_\rho h^{\tau\rho} ),
\eqno(4.3)$$
and
$$
 B= 2\int dt d\Omega r^{D-2} (-\delta h^{ra}\nabla _a h^b{}_b -\nabla_ah^{ab}\delta h^c{}_c +\delta h^c{}_c \nabla^r h^d{}_d
- \delta h^{cd}\nabla ^r h_{cd} +2 \delta h^{rb} \nabla _c h^{c}{}_b),
\eqno(4.4)$$
where we have written the last term in tangent space.
\par
The boundary term in tangent space  is of the generic form $\int dt d\Omega r^{D-2}\delta h\nabla h$ and it will vanish if $h^{ab}$ has the asymptotic form
$$
h_{ab}(t,r,\theta^m) = {h^{(1)}_{ab}( \theta^m)\over r^{{D-2\over 2}}} +O(r^{-{D\over 2}}) , \ \
\xi^a(t,r,\theta^m)={  \xi^{(0)} {}^{a}( \theta^m)\over r^{{D-4\over 2}}} +O(r^{-{D-2\over 2}}).
\eqno(4.5)$$
It is straightforward to find the asymptotic behaviour for the fields with world indices, for example $h_{mn}\sim r^{-{(D-6)\over 2}}$. The leading terms are independent of time as a constraint arising from Poincare symmetry. We will explain this in detail in section five.
\par
Reading off the symplectic potential, we find that the conserved charges are
$$
Q_{\xi^0}= 2\int d\Omega r^{D-2}(-\nabla^r \xi^0 (h^\mu{}_\mu -2 h^0{}_0) +2 \xi^0\nabla^r h^m{}_m -2 \xi^0\nabla_m h^m{}^r -2 h^r{}_0\nabla^0 \xi^0)\,,
\eqno(4.6)$$
$$
Q_{\xi^r}= 2\int d\Omega  r^{D-2}(\nabla^0 \xi^r (h^\mu{}_\mu -2h^r{}_r) -2 \xi^r\nabla^0 h^m{}_m +2 \xi^r\nabla_m h^m{}^0 +2h^0{}_r\nabla^r\xi^r)\,,
\eqno(4.7)$$
$$
Q_{\xi^m}= 2\int d\Omega  r^{D-2}(-2 h^r{}_m \nabla^0 \xi^m +2 \xi^m(\nabla^0h_m{}^r-\nabla^r h^0{}_m) +2 h^0{}_m \nabla^r \xi^m )\,.
\eqno(4.8)$$
The sum over $\mu$ in these equations  is over $0,r,m$.
\par
Using equation (4.5) we find that the charges formulated with the metric and parameter in tangent space go as  $ r^{D-2}r^{-{D-2\over 2}}r^{-1}r^{-{D-4\over 2}}\sim r^0$  if we take the leading terms. There is one exception to this for the terms that contain the time derivatives of the leading term which vanishes and so for these terms we find the sub-leading terms. The charges are given by
$$
Q_{\xi^0}= -2\int d\Omega (\xi^{(0)0}(2\nabla_m h^{(1)m}{}_r+{D-4\over 2}(h^{(1)t}{}_t-h^{(1)r}{}_r +h^{(1)m}{}_m)
+ (D-2) h^{(1)} {}^r{}_r)
$$
$$
+2h^{(1)0r}\partial_0 \xi^{(1)0})\,,
\eqno(4.9)$$
$$
Q_{\xi^r}= -2\int d\Omega  (\xi^{(0)r} (2\nabla_m h^{(1)m}{}_0
-2\partial_0 h^{(2)m}{}_m +D h_{0r}^{(1)}
-\partial_0 \xi^{(1)r}(h^{(1)t}{}_t- h^{(1)r}{}_r+h^{(1)m}{}_m))\,,
\eqno(4.10)$$
$$
Q_{\xi^m}= -4\int d\Omega  (\xi^{(0)m}(h^{(1)}_{0m}+\partial_0 h^{(2)}_{rm})-\partial_0 \xi^{(1)m} h^{(1)}_{rm}  )\,.
\eqno(4.11)$$
In these last three equations the covariant derivative is only with respect to the curved coordinates on the sphere. Note that the subleading components $\xi^{(1)a}$ appear through their time derivative, yielding independent charges in addition to those associated with the leading components $\xi^{(0)a}$. In [26-27] gravitational charges associated with similar subleading diffeomorphisms have also been obtained. It would be interesting to further investigate their relations to those presented above. 
\par
The Lie algebra of asymptotic symmetries parametrised by the four functions $\xi^{(0)a}(\theta^m)$ on the sphere is induced by the standard Lie bracket of vector fields $[\xi_1, \xi_2]^\mu=\xi_1^\nu \partial_\nu \xi_2^\mu-\xi_2^\nu \partial_\nu \xi_1^\mu$ and turns out to be abelian,
$$
\left[\xi_1^{(0)}(\theta^m),\xi_2^{(0)}(\theta ^m)\right]=0\,.
\eqno(4.12)$$
To see this we evaluate $[\xi_1, \xi_2]^\mu$ using the asymptotic expansion for the parameter of equation (4.5) but converted to world indices. One finds that  the composite parameter contains an additional factor of $r^{-1}$ and so can be discarded  as $r\to \infty$. In the appendix we show that BMS supertranslations are contained within this abelian algebra. However we find more asymptotic symmetries as a result of not fixing a gauge.
\par
The Poisson algebra of charges should be identically zero  up to central terms, i.e.,
$$
\{Q[\xi_1^{(0)}]\,, Q[\xi_2^{(0)}] \}=C(\xi_1^{(0)},\xi_2^{(0)})\,,
\eqno(4.13)$$
where the function $C(\xi_1^{(0)},\xi_2^{(0)})$ is field-independent. We find that this is automatically the case in the present context of linearised gravity since the charges are linear in the graviton field.
Setting $\xi^{(1)a}=0$ for simplicity, the expression for $C(\xi_1^{(0)},\xi_2^{(0)})$
is given by
$$
C(\xi_1^{(0)},\xi_2^{(0)})=-4 \int d\Omega\, (\xi_1^{(0)0} \nabla^2 \xi_2^{(0)r}-{D(D-4)\over 4}\xi_1^{(0)0}\xi_2^{(0)r}-\xi^{(0)0}_1 \nabla_m \xi_2^{(0)m} -(1 \leftrightarrow 2))\,.
\eqno(4.14)
$$
For generic (spi)-supertranslations in $D=4$ this central extension vanishes identically, since in this case we have $\xi^{(0)m}=\gamma^{mn}\partial_n \xi^{(0)r}$ (see appendix).
\par
We will now discuss the  parent action [62-64]
$$
{1\over 2}\int d^D x \{\epsilon^{\mu \nu\ldots \tau_{D-2}}
Y_{\tau_1\ldots \tau_{D-2}}{}^\rho C_{\mu\nu,\rho}
+(-{1\over 2} {(D-3)\over (D-2)}
Y_{\tau_1\ldots \tau_{D-2},}{}^{\rho}
Y^{\tau_1\ldots \tau_{D-2},}{}_{\rho}
$$
$$+{1\over 2}  (D-2) Y_{\tau_1\ldots\tau_{D-3}\rho,}{}^{\rho}
Y^{\tau_1\ldots\tau_{D-2}\kappa ,}{}_{\kappa}
-{1\over 2}Y_{\tau_1\ldots\tau_{D-3}\kappa,}{}_{\rho}
Y^{\tau_1\ldots\tau_{D-3} \rho,}{}^{\kappa})\sqrt {-\det g}\}\,,
\eqno(4.15)$$
where $ C_{\mu\nu ,}{}^\rho= \nabla _\mu h_\nu{}^\rho -\nabla _\nu h_\mu{}^\rho $,  $Y_{\tau_1\ldots \tau_{D-2},}{}^\rho $ is an independent field and  the gravity field $h_{\mu\nu}$ is not symmetric.  The  equations of motion are
$$
 \epsilon^{\mu \tau_1\ldots \tau_{D-1}}\nabla_{\tau_1}
Y_{\tau_2\ldots \tau_{D-1}}{}^\rho =0\,,
\eqno(4.16)$$
$$
 \epsilon_{\mu\nu}{}^{\tau_1\ldots \tau_{D-2}}
Y_{\tau_1\ldots \tau_{D-2},\rho}= -C_{\mu\nu, \rho}+C_{\nu \rho, \mu}
-C_{\mu \rho, \nu}+2(\eta_{\nu \rho}C_{\mu \lambda}{}^\lambda -\eta_{\mu \rho}C_{\nu \lambda}{}^\lambda )\,.
\eqno(4.17)$$
\par
The first of these equations implies that
$$ Y_{\tau_1\ldots \tau_{D-2},b}=\nabla _{[ \tau_1}
h_{\tau_2\ldots \tau_{D-2}], b}\,.
\eqno(4.18)$$
Substituting this back into the action of equation (4.15) we find the action for gravity in terms of its dual gravity field on the other hand eliminating $Y_{\tau_1\ldots \tau_{D-2},}{}^\rho $ we find the usual action for linearised gravity.
\par
The action of equation (4.15) is invariant under the gauge transformations
$$
\delta h_{a_1 \ldots a_{D-3},b}= \partial_{[b}\Lambda_{a_1\ldots a_{D-3}]} ,\ \  \delta h_{ab} = \nabla^c \Lambda_{cab},\ \ {\rm where}\ \  \epsilon^{e_1e_2e_3a_1\ldots a_{D-3}}\Lambda_{a_1\ldots a_{D-3}}\equiv 2\Lambda^{e_1e_2e_3}\,,
\eqno(4.19)$$
as well as the obvious invariance
$$
\delta h_{a_1 a_2\ldots a_{D-3},b}= \partial_{[a_1}\Lambda_{a_2\ldots a_{D-3}],b} ,\ \ \delta h_{ab}=0\,.
\eqno(4.20)$$
\par
Varying the action of equation (4.7) we find the equations of motion and the boundary term
$$
\int d t d\Omega r^{D-2} \epsilon ^{0bc_1\ldots c_{D-2}} Y_{c_1\ldots c_{D-2}, }{}^e \delta h_{be}\,.
\eqno(4.21)$$
Demanding that this term vanish as $r\to \infty$  requires that $h_{ab}$ has the asymptotic expansion of equation (4.5) and
$Y_{a_1\ldots a_{D-2}}{}^b $ goes as $r^{-{D\over 2}}$.
\par
The symplectic potential is given by
$$
 \theta [\delta h_{\nu\rho} ]= \int d^{D-2}\theta dr \epsilon^{0\nu \tau_1\ldots \tau_{D-2} } Y_{\tau_1\ldots \tau_{D-2} }{}^\rho \delta h_{\nu\rho}\,.
\eqno(4.22)$$
Using this we find that  the conserved charges are given by
$$
Q= \int d\Omega r^{{D-2}}\{-\epsilon ^{0r a_1\ldots a_{D-2} } Y_{a_1\ldots a_{D-2} ,}{}^c \xi_c
-h_{a 0}\nabla_c \Lambda^{c r a } +2 h^c{}_c \nabla_a \Lambda ^{0 a r}
-2 h_c {} ^r \nabla_a \Lambda ^{0 a c}\}\,.
\eqno(4.23)$$
There charge corresponding to the gauge transformations of equation (4.20) vanishes.
\par
The action of equation (4.15) contains the antisymmetric part of the graviton and it is also  invariant under the transformations
$$
\delta Y_{a_1\ldots a _{D-2},b}= {1\over 2} \nabla _ {[a_1}\Lambda _{a_2\ldots a_{D-2}]} ,\ \ \delta h_{ab}= \Lambda _{ab}= - \Lambda _{ba}\,,
\eqno(4.24)$$
where the two parameters are related by
$\Lambda _{a_2\ldots a_{D-2}} = {(-1)^{D}\over 2 (D-3)!}\epsilon _{a_1\ldots a_{D-2} c_1c_2} \Lambda ^{c_1c_2}$.
The corresponding charge is
$$
Q= -2\int d\Omega (-\Lambda_{r\alpha} h^{\alpha}{}_0 +\Lambda _0{}^\alpha h_\alpha {}_r)\,.
\eqno(4.25)$$
We hope to discuss the physical meaning of the charges in equations (4.23) and (4.25) in a later paper.
\par


We will now consider an alternative parent action for gravity which is closer in spirit to the parent action that we considered for the spin one case. One advantage is that we will find the Taub-Nut charge. We start from the action 
$$
\int d^D x \sqrt {-\det g} \{-{\over 4} C_{ab, c} C^{ab, c} -{1\over 2}C_{ab, c}C^{ac, b} +C_{ac,}{}^{c}C_{ab,}{}^{b}
$$
$$
+\epsilon ^{c_1\ldots c_{D-2} e_1e_2} C_{e_1e_2,}{}^{ d} \nabla_{c_1} \tilde h _{c_2\ldots c_{D-2}, d}
\}
\eqno(4.26)$$
where now $C_{ab,c}$ is an independent field.  The equation of motion for $h _{c_1\ldots c_{D-3}, d}$  implies that 
$C_{\mu\nu, \rho}= \partial_\mu h_{\nu \rho} - \partial_\nu h_{\mu \rho} $ and substituting this back into the action we find the linearised Einstein theory. The equation of motion for $C_{ab,c}$ is given by 
$$
C_{ab, c}+C_{ac, b}-C_{bc,a} -4 \delta_{c[b} C_{a] e,}{}^{e} -4 \epsilon_{ab}{}^{e_1\ldots e_{D-2}}\nabla_{e_1}\tilde h_{e_2\ldots e_{D-2}, c}=0
\eqno(4.27)$$
Substituting this back into the action we find the action for the dual graviton. We may rewrite this equation as 
$$
\omega_{c,ab}+2\omega^{e}{}_{,e [a}\delta _{b], c} +\epsilon ^{abe_1\ldots e_{D-2}} \nabla_{e_1}\tilde h_{e_2\ldots e_{D-2}, c}=0
\eqno(4.28)$$
where $\omega_{a,bc}$ is the linearised spin connection which is given by 
$$
\omega_{c.ab}= -{1\over 2} (C_{ab, c} +C_{ac,b}-C_{bc, a} )
\eqno(4.29)$$
\par
The boundary term for the parent action of equation (4.26) is given by 
$$
 \int dt d\Omega r^{D-2} \epsilon ^{r e_1\ldots e_{D-2} b_1b_2} C_{b_1b_2, }{}^d \delta \tilde h_{e_1\ldots e_{D-2}, d}
\eqno(4.30)$$
This will vanish if the fields have the fall off 
$$
h_{ab}(t,r,\theta^m) = {h^{(1)}_{ab}( \theta^m)\over r^{{D-2\over 2}}} +O(r^{-{D\over 2}}) , \ \
\tilde h_{a_1\ldots a_{D-3}, b}(t,r,\theta^m) = {\tilde h^{(1)}_{a_1\ldots a_{D-3}}( \theta^m)\over r^{{D-2\over 2}}} +O(r^{-{D\over 2}}) 
\eqno(4.31)$$
This fall off is consistent with the equation of motion (4.27) which involves both fields. 
\par
The parent action of equation (1) has the symmetry 
$$
\delta \tilde h_{a_1\ldots a_{D-3}, b}= \nabla_{[a_1}\tilde \Lambda _{a_2\ldots a_{D-2}], b},\ \  \delta C_{ab,c}=0
\eqno(4.32)$$
The  parameter has the fall off 
$$
\tilde \Lambda _{a_1\ldots a_{D-3}}(t,r,\theta^m)={ \tilde  \Lambda _{a_1\ldots a_{D-3}}^{(0)} {}^{a}( \theta^m)\over r^{{D-4\over 2}}} +O(r^{-{D-2\over 2}}).
\eqno(4.33)$$
\par
The symplectic potential has the form 
$$
\theta [\delta \tilde h]=  \int dr d\Omega r^{D-2} \epsilon ^{0 e_1\ldots e_{D-2} b_1b_2} C_{b_1b_2, }{}^d \delta \tilde h_{e_1\ldots e_{D-2}, d}
\eqno(4.34)$$
Using equation (1.32) we find the infinite number of conserved charges corresponding to the symmetry of equation (4.33) are 
$$
Q_{\tilde \Lambda}= -\int d\Omega r^{D-2} \epsilon ^{ \alpha_1\ldots \alpha_{D-4} \beta_1\beta _2} C_{\beta_1\beta _2, }{}^d  \tilde \Lambda_{\alpha_1\ldots \alpha_{D-4}, d}
\eqno(4.35)$$
In four dimension the symmetry is $\delta \tilde h_{\mu\nu}= \partial_{\mu}\tilde \xi_\nu$ is given by 
$$
Q_{\tilde \Lambda}= 2\int d\Omega r^2 \epsilon^{\beta_1\beta_2} \partial_{\beta_1}  h_{\rho \beta_2}\tilde \xi^\rho 
\eqno(4.36)$$
For $\tilde \xi^\rho $ a constant this is the Taub-Nut charge discussed  in reference [78].  In that reference it was found by carrying out a duality rotation between the graviton and the dual graviton and then shown to give the  charge of the Taub-Nut solution. The extension of the Taub-Nut charge to be part of the infinite number of conserved asymptotic charges was discussed in the  papers [32,33,34] In this paper we have derived  for the first time the Taub-Nut charge and their asymptotic extension directly from the dual gravity field itself rather than by an indirect path. 


\medskip
{\bf  5.  A generic free theory }
\medskip
The reader will  have realised that the different theories we have studied all have a similar behaviour for the their asymptotic fall off and charges. In this section we will give a generic discussion for the asymptotic fall off at spatial infinity for any massless particle. We consider a  particle described by the field $A$  that has the free generic action
$$
S=-{1\over 2}\int d^D x \nabla A \nabla A \sqrt {-\det g}\,,
\eqno(5.1)$$
where $\nabla $ is the derivative. We have suppressed the indices  in this equation but the indices on the two factors of  $\nabla A$ can in principle be contracted in many possible ways. The actual form of the action is determined by the irreducible representation of the Poincare group that the particle belongs to  and the way it is embedded in the field $A$.
The action has the  gauge symmetry $\delta A= \nabla \Lambda$ associated with a massless particle. The reader will have no trouble restoring the indices in the equations in this section  for any given particle at lest up to equation (5.6).
\par
The variation of the action is of the form
$$
\delta S=\int d^D x \delta A( \nabla  \nabla A) \sqrt {-\det g}-\int d^D x  \partial (\delta A \nabla A \sqrt {-\det g} )\,.
\eqno(5.2)$$
 The boundary term can be written as
$$
-\int dt d\Omega r^{D-2}\delta A \nabla A\,,
\eqno(5.3)$$
where $d\Omega =d^{D-2}\theta \det \hat e$. The symplectic potential is given by
$$
 \theta[\delta A]=  \int dr d^{D-2}\theta \delta A \nabla A \sqrt {-\det g}\,,
\eqno(5.4)$$
Using equation (1.32) we conclude that
$$
\delta Q= \int dr d^{D-2}\theta \sqrt {-det g}( \delta A \nabla \nabla \Lambda - \nabla \Lambda \nabla \delta A)
\eqno(5.5)$$
where the gauge transformation of $A$ is $\delta A= \nabla \Lambda$. In all the cases we studied in this paper one can arrange the derivative so that this is an integral over the sphere at infinity and we believe that this is the case in general.  Taking this to be the  case the conserved charges  take the very generic form
$$
Q= -\int _{S^{D-2}}  d\Omega r^{D-2}(\Lambda \nabla A+ \nabla \Lambda A)
\eqno(5.6)$$
\par
Examining the boundary term of equation (5.6) we  find the asymptotic behaviour as $r\to \infty$  for the fields in tangent space is as follows
$$
A= {A^{(1)}(\theta^m) \over r^{{D-2\over 2}}} +O(r^{-{D\over 2}}) , \ \ F=  {F^{(2) } (\theta^m) \over r^{{D\over 2}}} +O(r^{-({D\over 2}+1)}) , \ \  \Lambda= {{\Lambda^{(0)}(\theta^m) }\over   r^{{(D-4)\over 2}}}+O( r^{-{(D-2)\over 2}}) ,
\eqno(5.7)$$
where $F= \nabla A $. In doing this we first deduce the fall off for the field and then  that for the field strength and gauge parameter using the fact that $\nabla $ increases the fall off by a power of $r^{-1}$.
\par
We will now discuss the constraints on the asymptotic expansion of the fields due to Poincare symmetry. We previously carried this out for the scalar field in section one, but here we will do this  from the tangent view point and so be able to apply it to all types of fields. Poincare transformations are a special type of diffeomorphisms that leave the flat metric invariant, but not necessarily the vierbeins. The effect of such a transformation  as well as a local Lorentz rotation on the vierbein is given by
$$
\delta e^a_\mu= \nabla_\mu\xi^\nu e^a_\nu -\Lambda^a_{\ b} e^b_\mu\,.
\eqno(5.8)$$
 By choosing a suitable Lorentz transformation we can choose this to vanish; the required choice is
$$
\Lambda_{ab}=e^\nu_a e^\mu_b \nabla_\mu \xi_\nu =  e^\nu_{[a} e^\mu_ {b ]} \nabla_\mu \xi_\nu  \,,
\eqno(5.9)$$
The last equations follows due to the Killing equation $\nabla_{(\mu} \xi_{\nu)}=0$. The advantage of this procedure is that we do not have to change the vierbein when carrying out a Poincare transformation.
\par
Let us carry out  a boost, see equation (1.10), which is parametrised by
$$
\xi_t= -r b\,, \qquad \xi_r=t b\,, \qquad \xi_m=r t\, \partial_m b\,,
\eqno(5.10)$$
The corresponding Local Lorentz rotation which leaves the veirbein given in the introduction in equations (0.3)  inert is given by
$$
\Lambda_{tr}=-b\,, \qquad \Lambda_{t\alpha}=-\hat e^m_\alpha \partial_m b\,, \qquad \Lambda_{r\alpha}=0\,, \qquad \Lambda_{\alpha \beta}=e^n_\alpha e^m_\beta \nabla_m \xi_n=O(r^{-1})\,.
\eqno(5.11)$$
These are either of order $O(r^0)$ and time-independent, or of order $O(r^{-1})$.
\par
Finally  we can study the transformation of dynamical fields under these boost transformations. For simplicity, we consider the case of a vector field $A_a$,  but the argument straightforwardly extends to any tensor field. Under the combined boost and local Lorentz transformation which leaves both the metric and the vierbein invariant, we have
$$
\delta A_a=\xi^\mu \partial_\mu A_a-\Lambda_a^{\ b} A_b\,,
\eqno(5.12)$$
where $\xi$ and $\Lambda$ are given above. In the above all components $A_a$ have been argued to satisfy the same falloff rate in $r$, and we have to ensure that these falloffs are preserved under Poincare transformations. There is one potentially problematic term which is given by
$$
\delta A_a=r b \partial_t A_a+...\,.
\eqno(5.13)$$
This term would violate the falloff conditions unless the leading order term in the large $r^{-1}$ expansion of $A_a$ is time-independent.
\par
An alternative way to arrive at the same conclusion is  as follows.  The equations of motion in tangent space place constraints on the tangent derivative $\nabla_a$ acting on the fields.  As we have mentioned the derivative $\nabla_a$ in tangent space   involves derivatives with respect to $r$ and $ \hat e_a{}^m\partial_m$ and when acting on any term these  lower its  fall off by a factor of $r$. However, the same is not true for the time derivative. Hence all the terms in the equation of motion
 involving derivatives other than time lower the fall off in $r$ and so the equations of motion set the time derivative of the leading term of the asymptotic expansion of the field  to zero.

\medskip
{\bf 6. Spin three halves}
\medskip
We will now show that there are an infinite number of conserved charges associated with the free spin ${3\over 2 }$ particle.
The free action in four dimensions  is given by
$$
-{i\over 2}\int d^4 x   \epsilon ^{\mu\nu\rho\kappa}  \bar  \psi_\mu \gamma_5\gamma_\nu \nabla_\rho\psi_\kappa\,,
\eqno(6.1)$$
where $\gamma_\mu=e_\mu{}^a \gamma_a$ and $\nabla_{\rho}$ contains the spin connection required for the coordinates on the sphere. The variation of the action is given by
$$
-i\int d^4 x   \epsilon ^{\mu\nu\rho\kappa} \delta  \bar  \psi_\mu \gamma_5\gamma_\nu D_\rho\psi_\kappa
-{i\over 2}\int d^4 x  \partial_\rho( \epsilon ^{\mu\nu\rho\kappa}  \bar  \psi_\mu \gamma_5\gamma_\nu \delta \psi_\kappa)\,.
\eqno(6.2)$$
The boundary term is given by
$$
-{i\over 2}\int dt d\Omega r^2  \epsilon ^{ra b c}  \bar  \psi_a \gamma_5\gamma_b \delta \psi_c\,,
\eqno(6.3)$$
while the symplectic potential is
$$
 \theta [\delta \psi_\kappa] = -{i\over 2} \int _\Sigma dr d^2\theta \epsilon ^{0\mu\rho\kappa}  \bar  \psi_\mu \gamma_5\gamma_\rho
 \delta \psi_\kappa\,.
\eqno(6.4)$$
Computing the conserved charge according to equation (1.32) we find that
$$
Q_{\epsilon} = -i \int d\Omega r^2 \theta \epsilon ^{\alpha \beta }  \bar  \epsilon \gamma_5\gamma_\alpha   \psi_\beta\,.
\eqno(6.5)$$
\par
If the leading behaviour of $\psi_a$ is $r^{-n}$ as $r\to \infty$  then demanding that the boundary term vanishes requires that
$n={3\over 2}$ and so $\epsilon $ goes as $r^{-{1\over 2}}$. As a result the charge of equation (6.5) is well defined and non-zero.
\par
We now briefly repeat the calculation in $D$ dimensions. The "spin ${3\over 2}$" action is
$$
-{1\over 2}\int d^D x \bar \psi_\mu \gamma^{\mu\nu\rho} \nabla_\nu \psi_\rho  \sqrt {-\det g}\,.
\eqno(6.6)$$
The conserved charges are
$$
Q_{\epsilon}= \int d\Omega r^{D-2} \bar \epsilon \gamma ^{0r}\gamma^\alpha \psi_\alpha\,.
\eqno(6.7)$$
As $r\to \infty$ the gravitino goes as $\psi _a\sim r^{-{D-1\over 2}}$ and  the supersymmetry parameter as $\epsilon  \sim r^{-{D-3\over 2}}$.
The conserved charges for the gravitino in four dimensions have been previously found in reference [47-49]. It would be interesting to reconcile this with the above expression but this may not be straight forward as the techniques used are very different.
\par
We can now list the conserved charges for the linearised eleven dimensional supergravity theory. The charges for the gravitino  , the  three form  are, respectively, given by
$$
Q_\epsilon= \int d\Omega  r^9 \bar \epsilon \gamma^{0r}\gamma ^\alpha\psi_\alpha ,\ \
Q_{A_3}= \int d \Omega  r^9\Lambda_{\alpha\beta} F^{0r \alpha\beta}  , \ \
\eqno(6.8)$$
while  the gravity charges being given in equations (4.6)-(4.8) taking $D=11$.
\par
Taking the parent action of equation (3.5) for the three form and the six form gauge fields we find the charge
$$
Q_{\Lambda _{\beta_1\ldots \beta_5}}= -{1\over 4!}{1\over 7!}\int d \Omega  r^9  \epsilon ^{\alpha_1\ldots \alpha_4 \beta_1\ldots \beta_5}\Lambda_{\beta_1\ldots \beta_5} F_{\alpha_1\ldots \alpha_4}\,.
\eqno(6.9)$$
\par
It would be interesting to investigate how the charges are transformed under supersymmetry. The gravity, two form and five  form charges should belong to the vector representation of $E_{11}$ but this also contains an infinite number of higher level charges [72]. As such one should expect to have infinite  number of higher level conserved asymptotic charges each of which has an infinite number of charges. We note that many of these charges correspond to the higher dual forms of the three form and graviton.

\medskip
{\bf 7. Spin one half}
\medskip
The action for a spin one half Majorana particle in four dimensions is
$$
-{1\over 2} \int d^4 x \bar \lambda \gamma^a (e^{-1})_a {}^\mu \nabla_\mu \lambda \sqrt {-\det g}\,.
\eqno(7.1)$$
The boundary term coming from the variation of the action is
$$
-{1\over 2} \int dt d\Omega r^2  \bar \lambda \gamma^r \delta  \lambda\,.
\eqno(7.2)$$
For this to vanish we require that $\lambda$ has the asymptotic behaviour
$$
\lambda (t,r,\theta^m ) = { \lambda ^{( {3\over 2})}(t, \theta^m) \over r^{{3\over 2}} }+ O( r^{-{5\over 2}})\,.
\eqno(7.3)$$
\par
The equation of motion is of the form
$$
(\gamma^t\partial_t+ \gamma^r\partial_r +{1\over r} \gamma^\alpha (\hat e^{-1})_\alpha {}^m\partial_m)\lambda=0\,.
\eqno(7.4)$$
Substituting in the above expansion we find that $\dot \lambda ^{( {3\over 2})}=0$ and so the quantity
$$
Q= \int d\Omega r^2 \bar \epsilon (\theta^m)  \lambda ^{( {3\over 2})} (\theta^m)\,,
\eqno(7.4)$$
is conserved for any choice of $\epsilon (\theta^m)$. The appearance of an infinite number of conserved quantities when there is no apparent gauge symmetry also occurred for the case of a spin zero particle which is represented by a scalar field and the derivation above closely parallels this case. It would be interesting to see if the above conserved charges can appear from some kind of dual representation of the spin one half particle as it did for the scalar.


\medskip
{\bf  Appendix : Supertranslations}
\medskip

In this appendix we want to connect with the standard supertranslations originally found by BMS at null infinity [6-8], but also more recently understood from the perspective of spatial infinity [17-18,23-25]. To this end we start by introducing the Beig-Schmidt coordinate system [73-74] which allows to connect spatial infinity with null infinity most easily by taking appropriate limits [18,75].

Outside the lightcone Minkowski space admits a foliation by three-dimensional hyperboloids with positive constant curvature (de Sitter space),
$$
\eta_{\mu\nu} dx^\mu dx^\nu=d\rho^2+ \rho^2 (-d\tau^2+\cosh^2 \tau\, \gamma_{mn} dx^m dx^n),
\eqno(A.1)$$
where $x^m$ are coordinates on the unit sphere and $\tau \in (-\infty,\infty)$ is the global de Sitter time. The transformation to spherical coordinates is then given by
$$
\tau =\tanh^{-1} \left({t \over r}\right)\,, \qquad \rho^2=r^2-t^2.
\eqno(A.2)$$
In Beig-Schmidt gauge fluctuations around the flat metric are commonly assumed to have the prescribed falloff behaviour [17]
$$
h_{\rho\rho}={2\sigma(x^c)\over \rho}+O(\rho^{-2})\,, \quad h_{\rho a}=O(\rho^{-2})\,, \quad h_{ab}=\rho f_{ab}(x^c)+O(\rho^0)\,,
\eqno(A.3)$$
where we collectively denote $x^a=(\tau,x^m)$. Following the discussion in [17], the limit $r \to \infty$ with $t/r \to 0$ is equivalent to the limit $\rho \to \infty$ and $\tau \to 0$, such that one finds the asymptotic behaviour
$$
h_{rr}={2\sigma(x^m) \over r}+O(r^{-2})\,, \quad
h_{tt}={f_{\tau\tau}(x^m) \over r}+O(r^{-2})\,, \quad
$$
$$
h_{tm}=f_{\tau m}(x^m)+O(r^{-1})\,, \quad
h_{mn}=r f_{mn}(x^m)+O(r^0)\,,
\eqno(A.4)$$
and
$$
h_{rt}=O(r^{-2})\,, \qquad
h_{rm}=O(r^{-1})\,.
\eqno(A.5)$$
The first set of falloffs in equation (A.4)  is in perfect agreement with those in equation (4.5) including the fact that  the leading functions are time-independent. However,  the  falloffs in equation (A.5) are actually stronger than what we imposed in section four, namely (4.5).
\par
In the usual approach the allowed diffeomorphisms are those that preserve the form of the metric that has been adopted, in this case that of equation (A.1), or equivalently in equations (A.4) and (A.5). In contrast the fall off of the field $h_{ab}$ is deduced from the requirement that the boundary terms in the variation of the action vanish. It also is the case that
the expression for the fall off of the parameter $\xi^a$ of equation (4.5)  preserves the form of the field $h_{ab}$ as is clear if one carries out  the diffeomorphism in tangent space. We will now examine what further restriction one must place on the parameter $\xi^a$ over and above those of equation (4.5) in order to recover the stronger fall offs in equation (4.5).
Taking the usual expression for a diffeomorphism of the the field $h_{\mu\nu}$,
$$
\delta_\xi h_{\mu\nu}=\xi^\lambda \partial_\lambda \eta_{\mu\nu}+\eta_{\mu \lambda} \partial_\nu \xi^\lambda+\eta_{\nu \lambda} \partial_\mu \xi^\lambda\,,
\eqno(A.6)$$
we find that
$$
\delta_\xi h_{rt}=r^{-1} \partial_t \xi^{r(0)}+O(r^{-2})\,, \quad \delta_\xi h_{rm}=\partial_m \xi^{r(0)}-\gamma_{mn} \xi^{n(0)}+O(r^{-1})\,.
\eqno(A.7)$$
For this to vanish in agreement with the fall off of equation (A.5), we need to impose that
$$
\partial_t \xi^{r(0)}=0\,, \qquad \xi^{m(0)}=\gamma^{mn} \partial_n \xi^{r(0)}\,.
\eqno(A.8)$$
In fact one can use $ \xi^{r(0)}$ and $ \xi^{r(0)}$ respectively to set the additional parts of $h_{rt}$ and $h_{rm}$ which are non-zero in this paper  to zero.
\par
We observe that the leading components $\xi^{m(0)}$ are fully determined in terms of $\xi^{r(0)}$. Thus it seems that we are left with two time-independent functions on the sphere, namely $\xi^{t(0)}$ and $\xi^{r(0)}$, while the well-known supertranslation symmetries are parametrised by a single such function. However the relation between symmetry parameters at spatial and null infinity is intricate. In Beig-Schmidt gauge, supertranslations are generated by the vector field [17]
$$
\xi^\rho=\omega+O(\rho^{-1})\,, \qquad \xi^\tau=-\rho^{-1}\, \partial_\tau \omega+O(\rho^{-2})\,, \qquad \xi^m=\rho^{-1}\, \gamma^{mn} \partial_n \omega+O(\rho^{-2})\,,
\eqno(A.9)$$
where $\omega=\omega(\tau,x^m)$ is any odd function under time-parity reversal $(\tau, x^m) \mapsto (-\tau, -x^m)$
which also satisfies the constraint $(D^2+3)\omega=0$ on the hyperboloid.
Using the change of coordinates  of equation (A.2) in the appropriate limit $\rho \to \infty$ and $\tau \to 0$, at leading order we find the relation
$$
\xi^\rho=\xi^{r(0)}+O(\rho^{-1})\,, \qquad \xi^\tau=\rho^{-1} \xi^{t(0)}+O(\rho^{-2})\,,
\eqno(A.10)$$
such that we can make the identification
$$
\xi^{r(0)}=\omega|_{\tau=0}\,, \qquad \xi^{t(0)}=-\partial_\tau \omega |_{\tau=0}\,.
\eqno(A.11)$$
Since $\omega$ is odd under time-parity reversal, the above expressions for $\xi^{r(0)}$ and $\xi^{t(0)}$ are odd and even under parity, respectively. Note that this result is in perfect agreement with the findings of Henneaux and Troessaert [23]. Thus supertranslations in spherical coordinates correspond to a choice of odd $\xi^{r(0)}$ and even $\xi^{t(0)}$ on the sphere. On the other hand different choices of parities for these functions correspond to more general spi-supertranslations [17-18,75-77].

\medskip
{\bf Ackowledgement}
\medskip
The work of KN is supported by the ERC Consolidator Grant N. 681908, ÒQuantum black holes: A microscopic window into the microstructure of gravityÓ. We would also like to thank the STFC, grant numbers ST/P000258/1   and ST/T000759/1, for support. We would like to thank Glenn Barnich for helpful discussions and comments on the paper.

\medskip
{\bf   References }
\medskip

[1] R.~L.~Arnowitt, S.~Deser and C.~W.~Misner,
``Dynamical Structure and Definition of Energy in General Relativity,''
Phys. Rev.{\bf  16} (1959), 1322-1330.

[2] A.~Komar,
``Covariant conservation laws in general relativity,''
Phys. Rev.{\bf  113} (1959), 934-936.

[3] T.~Regge and C.~Teitelboim,
``Role of Surface Integrals in the Hamiltonian Formulation of General Relativity,''
Annals Phys.{\bf  88} (1974), 286.

[4] L.~F.~Abbott and S.~Deser,
``Stability of Gravity with a Cosmological Constant,''
Nucl. Phys. B{\bf  195} (1982), 76-96.

[5] J.~D.~Brown and J.~W.~York, Jr.,
``Quasilocal energy and conserved charges derived from the gravitational action,''
Phys. Rev. D{\bf  47} (1993), 1407-1419
[arXiv:gr-qc/9209012 [gr-qc]].

[6] H.~Bondi, M.~G.~J.~van der Burg and A.~W.~K.~Metzner,
``Gravitational waves in general relativity. 7. Waves from axisymmetric isolated systems,''
Proc. Roy. Soc. Lond. A{\bf  269} (1962), 21-52.

[7] R.~K.~Sachs,
``Gravitational waves in general relativity. 8. Waves in asymptotically flat space-times,''
Proc. Roy. Soc. Lond. A{\bf  270} (1962), 103-126.

[8] R.~Sachs,
``Asymptotic symmetries in gravitational theory,''
Phys. Rev.{\bf  128} (1962), 2851-2864.

[9] J.~D.~Brown and M.~Henneaux,
``Central Charges in the Canonical Realization of Asymptotic Symmetries: An Example from Three-Dimensional Gravity,''
Commun. Math. Phys.{\bf  104} (1986), 207-226.

[10] R.~M.~Wald and A.~Zoupas,
``A General definition of 'conserved quantities' in general relativity and other theories of gravity,''
Phys. Rev. D{\bf  61} (2000), 084027
[arXiv:gr-qc/9911095 [gr-qc]].

[11] G.~Barnich and C.~Troessaert,
``BMS charge algebra,''
JHEP{\bf  12} (2011), 105
doi:10.1007/JHEP12(2011)105
[arXiv:1106.0213 [hep-th]].

[12] G.~Barnich and C.~Troessaert,
``Symmetries of asymptotically flat 4 dimensional spacetimes at null infinity revisited,''
Phys. Rev. Lett.{\bf  105} (2010), 111103
[arXiv:0909.2617 [gr-qc]].

[13] G.~Barnich and C.~Troessaert,
``Aspects of the BMS/CFT correspondence,''
JHEP{\bf  05} (2010), 062
doi:10.1007/JHEP05(2010)062
[arXiv:1001.1541 [hep-th]].

[14] G.~Barnich and P.~H.~Lambert,
``A Note on the Newman-Unti group and the BMS charge algebra in terms of Newman-Penrose coefficients,''
Adv. Math. Phys.{\bf  2012} (2012), 197385
[arXiv:1102.0589 [gr-qc]].

[15] G.~Barnich and C.~Troessaert,
``Finite BMS transformations,''
JHEP{\bf  03} (2016), 167
[arXiv:1601.04090 [gr-qc]].

[16] G.~Barnich, P.~Mao and R.~Ruzziconi,
``BMS current algebra in the context of the Newman-Penrose formalism,''
Class. Quant. Grav.{\bf  37} (2020) no.9, 095010 \break
[arXiv:1910.14588 [gr-qc]].

[17] G.~Compere and F.~Dehouck,
``Relaxing the Parity Conditions of Asymptotically Flat Gravity,''
Class. Quant. Grav.{\bf  28} (2011), 245016
[erratum: Class. Quant. Grav.{\bf  30} (2013), 039501]
[arXiv:1106.4045 [hep-th]].

[18] C.~Troessaert,
``The BMS4 algebra at spatial infinity,''
Class. Quant. Grav. {\bf 35} (2018) no.7, 074003
[arXiv:1704.06223 [hep-th]].

[19] \'E.~\'E.~Flanagan and D.~A.~Nichols,
``Conserved charges of the extended Bondi-Metzner-Sachs algebra,''
Phys. Rev. D{\bf  95} (2017) no.4, 044002
[arXiv:1510.03386 [hep-th]].

[20] M.~Campiglia and A.~Laddha,
``Asymptotic symmetries and subleading soft graviton theorem,''
Phys. Rev. D{\bf  90} (2014) no.12, 124028
[arXiv:1408.2228 [hep-th]].

[21] M.~Campiglia and A.~Laddha,
``New symmetries for the Gravitational S-matrix,''
JHEP{\bf  04} (2015), 076
doi:10.1007/JHEP04(2015)076
[arXiv:1502.02318 [hep-th]].

[22] M.~Campiglia and J.~Peraza,
``Generalized BMS charge algebra,''
Phys. Rev. D{\bf  101} (2020) no.10, 104039
[arXiv:2002.06691 [gr-qc]].

[23] M.~Henneaux and C.~Troessaert,
``BMS Group at Spatial Infinity: the Hamiltonian (ADM) approach,''
JHEP{\bf  03} (2018), 147
[arXiv:1801.03718 [gr-qc]].

[24] M.~Henneaux and C.~Troessaert,
``Hamiltonian structure and asymptotic symmetries of the Einstein-Maxwell system at spatial infinity,''
JHEP{\bf  07} (2018), 171\break
[arXiv:1805.11288 [gr-qc]].

[25] O.~Fuentealba, M.~Henneaux, S.~Majumdar, J.~Matulich and C.~Troessaert,
``Asymptotic structure of the Pauli-Fierz theory in four spacetime dimensions,''
Class. Quant. Grav.{\bf  37} (2020) no.23, 235011
[arXiv:2007.12721 [hep-th]]. 

[26] O.~Fuentealba, M.~Henneaux, J.~Matulich and C.~Troessaert,
``Bondi-Metzner-Sachs Group in Five Spacetime Dimensions,''
Phys. Rev. Lett. {\bf 128} (2022) no.5, 051103
[arXiv:2111.09664 [hep-th]].

[27] O.~Fuentealba, M.~Henneaux, J.~Matulich and C.~Troessaert,
``Asymptotic structure of the gravitational field in five spacetime dimensions: Hamiltonian analysis,''
JHEP {\bf 07} (2022), 149
[arXiv:2206.04972 [hep-th]].

[28] G.~Comp\`ere, A.~Fiorucci and R.~Ruzziconi,
``Superboost transitions, refraction memory and super-Lorentz charge algebra,''
JHEP{\bf  11} (2018), 200
[erratum: JHEP{\bf  04} (2020), 172]
[arXiv:1810.00377 [hep-th]].

[29] G.~Comp\`ere, A.~Fiorucci and R.~Ruzziconi,
``The $\Lambda$-BMS$_4$ charge algebra,''
JHEP{\bf  10} (2020), 205
[arXiv:2004.10769 [hep-th]].

[30] G.~Comp\`ere and D.~A.~Nichols,
``Classical and Quantized General-Relativistic Angular Momentum,''
[arXiv:2103.17103 [gr-qc]].

[31] L.~Freidel, R.~Oliveri, D.~Pranzetti and S.~Speziale,
``The Weyl BMS group and Einsteins equations,''
JHEP{\bf  07} (2021), 170
[arXiv:2104.05793 [hep-th]].

[32] H.~Godazgar, M.~Godazgar and C.~N.~Pope,
``New dual gravitational charges,''
Phys. Rev. D{\bf  99} (2019) no.2, 024013
[arXiv:1812.01641 [hep-th]].

[33] H.~Godazgar, M.~Godazgar and C.~N.~Pope,
``Tower of subleading dual BMS charges,''
JHEP{\bf  03} (2019), 057
[arXiv:1812.06935 [hep-th]].

[34] H.~Godazgar, M.~Godazgar and M.~J.~Perry,
``Asymptotic gravitational charges,''
Phys. Rev. Lett.{\bf  125} (2020) no.10, 101301
[arXiv:2007.01257 [hep-th]].

[35] M.~Geiller and C.~Zwikel,
``The partial Bondi gauge: Further enlarging the asymptotic structure of gravity,''
[arXiv:2205.11401 [hep-th]].


[36] T.~He, P.~Mitra, A.~P.~Porfyriadis and A.~Strominger,
``New Symmetries of Massless QED,''
JHEP{\bf  10} (2014), 112
[arXiv:1407.3789 [hep-th]].

[37] M.~Campiglia and R.~Eyheralde,
``Asymptotic $U(1)$ charges at spatial infinity,''
JHEP{\bf  11} (2017), 168
[arXiv:1703.07884 [hep-th]].

[38] M.~Campiglia and A.~Laddha,
``Asymptotic charges in massless QED revisited: A view from Spatial Infinity,''
JHEP{\bf  05} (2019), 207
[arXiv:1810.04619 [hep-th]].

[39] V.~Hosseinzadeh, A.~Seraj and M.~M.~Sheikh-Jabbari,
``Soft Charges and Electric-Magnetic Duality,''
JHEP{\bf  08} (2018), 102
[arXiv:1806.01901 [hep-th]].

[40] M.~Henneaux and C.~Troessaert,
``Asymptotic symmetries of electromagnetism at spatial infinity,''
JHEP{\bf  05} (2018), 137
[arXiv:1803.10194 [hep-th]].

[41] M.~Henneaux and C.~Troessaert,
``Asymptotic structure of electromagnetism in higher spacetime dimensions,''
Phys. Rev. D {\bf 99} (2019) no.12, 125006
[arXiv:1903.04437 [hep-th]].

[42] M.~Henneaux and C.~Troessaert,
``A note on electric-magnetic duality and soft charges,''
JHEP{\bf  06} (2020), 081
[arXiv:2004.05668 [hep-th]].

[43] L.~Freidel and D.~Pranzetti,
``Electromagnetic duality and central charge,''
Phys. Rev. D{\bf  98} (2018) no.11, 116008
[arXiv:1806.03161 [hep-th]].


[44] A.~Strominger,
``Asymptotic Symmetries of Yang-Mills Theory,''
JHEP{\bf  07} (2014), 151
[arXiv:1308.0589 [hep-th]].

[45] G.~Barnich and P.~H.~Lambert,
``Einstein-Yang-Mills theory: Asymptotic symmetries,''
Phys. Rev. D{\bf  88} (2013), 103006
[arXiv:1310.2698 [hep-th]].


[46] T.~T.~Dumitrescu, T.~He, P.~Mitra and A.~Strominger,
``Infinite-dimensional fermionic symmetry in supersymmetric gauge theories,''
JHEP{\bf  08} (2021), 051 \break
[arXiv:1511.07429 [hep-th]].

[47] V.~Lysov,
``Asymptotic Fermionic Symmetry From Soft Gravitino Theorem,'' \break
[arXiv:1512.03015 [hep-th]].

[48] S.~G.~Avery and B.~U.~W.~Schwab,
``Residual Local Supersymmetry and the Soft Gravitino,''
Phys. Rev. Lett.{\bf  116} (2016) no.17, 171601
[arXiv:1512.02657 [hep-th]].

[49] O.~Fuentealba, M.~Henneaux, S.~Majumdar, J.~Matulich and T.~Neogi,
``Asymptotic structure of the Rarita-Schwinger theory in four spacetime dimensions at spatial infinity,''
JHEP {\bf 02} (2021), 031
[arXiv:2011.04669 [hep-th]].

[50] M.~Henneaux, J.~Matulich and T.~Neogi,
``Asymptotic realization of the super-BMS algebra at spatial infinity,''
Phys. Rev. D {\bf 101} (2020) no.12, 126016
[arXiv:2004.07299 [hep-th]].

[51] O.~Fuentealba, M.~Henneaux, S.~Majumdar, J.~Matulich and T.~Neogi,
``Local supersymmetry and the square roots of Bondi-Metzner-Sachs supertranslations,''
Phys. Rev. D {\bf 104} (2021) no.12, L121702
[arXiv:2108.07825 [hep-th]].


[52] A.~Campoleoni, D.~Francia and C.~Heissenberg,
``On higher-spin supertranslations and superrotations,''
JHEP {\bf 05} (2017), 120
[arXiv:1703.01351 [hep-th]].

[53] A.~Campoleoni, D.~Francia and C.~Heissenberg,
``Asymptotic Charges at Null Infinity in Any Dimension,''
Universe {\bf 4} (2018) no.3, 47
[arXiv:1712.09591 [hep-th]].

[54] A.~Campoleoni, D.~Francia and C.~Heissenberg,
``On asymptotic symmetries in higher dimensions for any spin,''
JHEP {\bf 12} (2020), 129
[arXiv:2011.04420 [hep-th]].


[55] A.~Strominger,
``On BMS Invariance of Gravitational Scattering,''
JHEP{\bf  07}  \break (2014),  152
doi:10.1007/JHEP07(2014)152
[arXiv:1312.2229 [hep-th]].

[56] T.~He, V.~Lysov, P.~Mitra and A.~Strominger,
``BMS supertranslations and Weinbergs soft graviton theorem,''
JHEP{\bf  05} (2015), 151
doi:10.1007/JHEP05(2015)151
[arXiv:1401.7026 [hep-th]].

[57] S.~Weinberg,
``Infrared photons and gravitons,''
Phys. Rev.{\bf  140} (1965), B516-B524.

[58] A.~Strominger,
``Lectures on the Infrared Structure of Gravity and Gauge Theory,''
[arXiv:1703.05448 [hep-th]].


[59] G.~Barnich and F.~Brandt,
``Covariant theory of asymptotic symmetries, conservation laws and central charges,''
Nucl. Phys. B {\bf 633} (2002), 3-82
[arXiv:hep-th/0111246 [hep-th]].

[60] G.~Barnich and G.~Compere,
``Surface charge algebra in gauge theories and thermodynamic integrability,''
J. Math. Phys. {\bf 49} (2008), 042901
[arXiv:0708.2378 [gr-qc]].


[61] A.~Strominger,
``Magnetic Corrections to the Soft Photon Theorem,''
Phys. Rev. Lett.{\bf  116} (2016) no.3, 031602
[arXiv:1509.00543 [hep-th]].

[62]  P. West, ``E(11) and M theory,'' Class. Quant. Grav. 18 (2001) 4443-4460,
arXiv:hep-th/0104081 [hep-th].

[63] P.~West, ``Very extended E(8) and A(8) at low levels, gravity and supergravity,''
Class. Quant. Grav. 20 (2003) 2393?2406, arXiv:hep-th/0212291 [hep-th].

[64] N. Boulanger, P. P. Cook, and D. Ponomarev, ``Off-Shell Hodge Dualities in Linearised Gravity and E11,'' JHEP 09 (2012) 089, arXiv:1205.2277 [hep-th].

[65] N. Boulanger, Per Sundell and P. West, {\it Gauge fields and infinite chains of dualities}, JHEP 1509 (2015) 192,  arXiv:1502.07909.


[66] M.~Campiglia, L.~Coito and S.~Mizera,
``Can scalars have asymptotic symmetries?,''
Phys. Rev. D{\bf  97} (2018) no.4, 046002
[arXiv:1703.07885 [hep-th]].

[67] M.~Campiglia, L.~Freidel, F.~Hopfmueller and R.~M.~Soni,
``Scalar Asymptotic Charges and Dual Large Gauge Transformations,''
JHEP{\bf  04} (2019), 003
[arXiv:1810.04213 [hep-th]].

[68] D.~Francia and C.~Heissenberg,
``Two-Form Asymptotic Symmetries and Scalar Soft Theorems,''
Phys. Rev. D {\bf 98} (2018) no.10, 105003
[arXiv:1810.05634 [hep-th]].

[69] A.~Seraj,
``Gravitational breathing memory and dual symmetries,''
JHEP {\bf 05} (2021), 283
[arXiv:2103.12185 [hep-th]].

[70] M.~Henneaux and C.~Troessaert,
``Asymptotic structure of a massless scalar field and its dual two-form field at spatial infinity,''
JHEP{\bf  05} (2019), 147
[arXiv:1812.07445 [hep-th]].

[71] H.~Afshar, E.~Esmaeili and M.~M.~Sheikh-Jabbari,
``Asymptotic Symmetries in $p$-Form Theories,''
JHEP {\bf 05} (2018), 042
[arXiv:1801.07752 [hep-th]].

[72] See the review P. West,
{\it A brief review of E theory}, Proceedings of Abdus Salam's 90th  Birthday meeting, 25-28 January 2016, NTU, Singapore, Editors L. Brink, M. Duff and K. Phua, World Scientific Publishing and IJMPA, {\bf Vol 31}, No 26 (2016) 1630043, arXiv:1609.06863.


[73] R.~Beig and B.G.~Schmidt,
``Einstein's equations near spatial infinity'', Comm. Math. Phys. 87 (1982) 65.

[74] R.~Beig,
``Integration of Einstein's equations near spatial infinity'', Proc. R. Soc. A 391 (1984) 295.

[75] F.~Capone, K.~Nguyen and E.~Parisini,
``Charge and Antipodal Matching across Spatial Infinity,''
[arXiv:2204.06571 [hep-th]].

[76] A.~Ashtekar and R.~O.~Hansen, ``A unified treatment of null and spatial infinity in general relativity. I - Universal structure, asymptotic symmetries, and conserved quantities at spatial infinity,'' J. Math. Phys. {\bf 19} (1978), 1542-1566.

[77] A.~Ashtekar and A.~Magnon, ``From i0 to the 3+1 description of spatial infinity'', J.Math.Phys. {\bf 25} (1984) 2682.

[78] R.  Argurio, F. Dehouck and L. Houart,  ``Supersymmetry and Gravitational Duality'',  Phys.Rev. {\bf D 79} (2009) 125001, arXiv:0810.4999.

\end